\documentclass[12pt]{article}
\usepackage{amsmath,amsfonts,amssymb,amscd,amsxtra,amsthm}

\usepackage{graphicx}
\usepackage{epsfig}
\usepackage{bm}

\begin{document}
\title{Leptonic decay of Heavy-light Mesons in a QCD Potential Model }
\author{ $^{1}$Krishna Kingkar Pathak, $^{2}$D K Choudhury and $^{3}$N S Bordoloi \\
$^{1}$Deptt. of Physics,Arya Vidyapeeth College,Guwahati-781016,India\\
e-mail:kkingkar@gmail.com\\
$^{2}$Deptt. of Physics, Gauhati University, Guwahati-781014,India\\
$^{3}$Deptt. of Physics, Cotton College, Guwahati-781001}
\date{}
\maketitle
\begin{abstract}
  We study the masses and decay constants of heavy-light flavour  mesons $D$, $D_{s}$, $B$ and $B_{s}$ in a QCD Potential model . The mesonic wavefunction is used to compute the masses of $D$ and $B$ mesons in the ground state and the wavefunction is transformed to momentum space to estimate the pseudoscalar decay constants of these mesons. The leptonic decay widths and branching ratio of these mesons for different leptonic channels are also computed to compare with the experimental values. The results are found to be compatible with available data.   
\end{abstract}

\section{Introduction}
Heavy hadron spectroscopy has played a major role in the foundation of QCD. In the last few years however it has sparked a renewal of interest due to the numerous data available from the B factories, CLEO, LHCb,the Tevatron and by the progress made in the theoretical methods. The remarkable progress at the experimental side, for the study of hadrons has opened up new challenges in the theoretical understanding of light-heavy flavour hadrons.\\

The study of the wave functions of heavy-flavored mesons like B and D are important both analytically and numerically  for studying the properties of strong interaction between heavy and light quarks as well as for investigating the mechanism of heavy meson decays. The wave function determines the momentum distributions of the quark and antiquark in mesons, which is an important quantity for calculating the amplitude,formfactors and decay constants of heavy meson decays [1-7]. The Pseudoscalar decay constants of $B$ and $D$ mesons which are related to the wave-function overlap of the quark and antiquark is an important parameter for the determination of the Cabibbo-Kobayashi-Maskawa (CKM) quark-mixing matrix element $|V_{ub}|$ or $|V_{cd}|$. If the CKM element is well measured then one can compute the decay constants from the experimental decay rate as well. Following the same path, the theoretical input on Pseudoscalar decay constants
can allow a determination of the CKM element which provides a direct test of the Standard Model.\\
 Regarding the value of the Pseudoscalar decay constants, experiments and lQCD calculations agree very well with each other for D meson. But for $D_{s}$ meson noticable discrepancy is seen between the  PDG average[8] and Lattice QCD values[9]. As argued in ref.[10],this discrepancy is very interesting since the same systematic error effects the lattice calculation of $f_{D}$ and $f_{D_{S}}$. In the later part, however this discrepancy is reduced to certain extent by the updated data from the experiments[11-14], lattest PDG values[15] and the updated results of HPQCD Collaboration[16]. Again $f_{B_{s}}$ cannot be measured experimentally due to its neutrality in nature, hence it has to be determined from theory.\\
 Based on its quark structure of meson there exist many Potential Models in the literature [17-19], with the QCD potential considering the combination coulomb term and linear confining term.  The present authors have been persuing a specific potential model with  $V(r)=\frac{-4\alpha_{s}}{3r}+br+c$ [20-22]  considering its Coulombic part as perturbation[23] and linear as parent as well as linear part as perturbation [22,24,25] and coulombic part as parent.
	In this work, we have used the wavefunction with  linear part as perturbation and transformed it to momomentum space by applying Fourier transformation. This wavefunction is then used to study decay constants and leptonic branching ratio of $B$ and $D$ mesons in this QCD Potential model approach. \\
  
 We discuss the formalism in section 2 and summarise the results and conclusion in section 3.\\
	        
\section{Formalism}
\subsection{Wave function in the model}
The non relativistic predictions of Potential Models with a nonrelativistic Hamiltonian for the heavy-light and heavy-heavy mesons are found to be in fair agreements with the updated theoretical, experimental and Lattice results[23-28]. Hence, we start with the ground state ($l=0$) spin independent non relativistic Fermi-Breit Hamiltonian (without the contact term)
\begin{equation}
   H=-\frac{\nabla^{2}}{2\mu}-\frac{4\alpha_s}{3r}+br+c. 
\end{equation}
With the linear term $br+c$ as perturbation and using Dalgarno method,the wave function in the model is obtained as [24,25] :
\begin{equation}
\psi_{rel+conf}\left(r\right)=\frac{N^{\prime}}{\sqrt{\pi a_{0}^{3}}} e^{\frac{-r}{a_{0}}}\left( C^{\prime}-\frac{\mu b a_{0} r^{2}}{2}\right)\left(\frac{r}{a_{0}}\right)^{-\epsilon}
\end{equation}
\begin{equation}
N^{\prime}=\frac{2^{\frac{1}{2}}}{\sqrt{\left(2^{2\epsilon} \Gamma\left(3-2\epsilon\right) C^{\prime 2}-\frac{1}{4}\mu b a_{0}^{3}\Gamma\left(5-2\epsilon\right)C^{\prime}+\frac{1}{64}\mu^{2} b^{2} a_{0}^{6}\Gamma\left(7-2\epsilon\right)\right)}}
\end{equation}
\begin{equation}
C^{\prime}=1+cA_{0}\sqrt{\pi a_{0}^{3}}
\end{equation}
\begin{equation}
\mu=\frac{m_{i}m_{j}}{m_{i}+m_{j}}
\end{equation}
\begin{equation}
a_{0}=\left(\frac{4}{3}\mu \alpha_{s}\right)^{-1}
\end{equation}
\begin{equation}
\epsilon=1-\sqrt{1-\left(\frac{4}{3}\alpha_{s}\right)^{2}}.
\end{equation}
The QCD potential is taken as
\begin{equation}
V\left(r\right)=-\frac{4}{3r}\alpha_{s}+br+c
\end{equation}
Here $A_{0}$ is the undetermined factor appearing in the series solution of the Schr\"odinger equation(Eq.A.39 in the Appendix).                                                                                                                                                                                                                       The term $\left(\frac{r}{a_{0}}\right)^{-\epsilon}$ in eq.2 is the Dirac factor and was introduced to incorporate relativistic effect [22,24,25].  \\
The wavefunction in momentum space can be obtained by using the Fourier transform as
\begin{equation}
\psi\left(p\right)=\frac{1}{\left(2\pi \hbar c\right)^{3/2}}\int d^3re^{\frac{-i\overline{p}.\overline{r}}{\hbar c}}\psi\left(r\right).
\end{equation}
Separating the variable-dependence of the momentum space wave function as[7]
\begin{equation}
\psi\left(\overline{p}\right)=\psi_{l}\left(p\right)Y_{lm}\left(\theta,\phi\right)
\end{equation}
one can obtain for $l=0$ in the natural unit as[16]
\begin{equation}
\psi\left(p\right)=\sqrt{\frac{2}{\left(\pi p^2 \right)}}\int dr sin\left(pr\right)\psi(r).
\end{equation}
Then using Eqs.(2)and(11) and the standard result  
\begin{equation}
\int x^{p-1}e^{-ax}sin(mx)dx=\frac{\Gamma(p) sin(p\theta)}{(a^2+m^2)^{1/2}},
\end{equation}
one can obtain the normalised wavefunction in momentum space as
\begin{equation}
\psi_{rel+conf}\left(p\right)=\frac{N\sqrt{2}\left(2-\epsilon\right)\Gamma\left(2-\epsilon\right)}{\pi\left(1+ a_{0}^{2}p^{2}\right)^{\frac{3-\epsilon}{2}}}\left[C^{\prime}-\frac{\left(4-\epsilon\right)\left(3-\epsilon\right)\mu b a_{0}^{3}}{2\left(1+a_{0}^{2}p^{2}\right)}\right].
\end{equation}
This simplified form of the wavefunction  gives the momentum distribution of the quark and antiquark.

\subsection{Masses and Decay constants of $D$ and $B$ mesons}
 
 The decay constant with relativistic correction can be expressed
through the meson wave function $\psi_{P}(p)$ in the momentum
space [29,30]as\\
\begin{equation}
f_{P} =\sqrt{\frac{12}{M_{P}}}\int\frac{d^3p}{(2\pi)^3} \left(
\frac{E_q + m_q}{2E_q}\right)^{1/2} \left(\frac{E_{\bar
q} + m_{\bar q}}{2E_{\bar q}}\right)^{1/2}\left(1+
\lambda_{P} \frac{p^2}{[E_q+m_q][E_{\bar q}+m_{\bar
q}]}\right)\psi_{P}(p)
\end{equation}
 with $\lambda_P=-1$ for pseudoscalar mesons and $E_{q}=\sqrt{p^{2}+m_{q}^{2}}$.\\
The pseudoscalar mass $M_{P}$ of the mesons are calculated by using the relation[27,31]
\begin{equation} 
M_{p}=m_{Q}+m_{\overline Q}+
\langle H \rangle
\end{equation}
where the expectation value of the hamiltonian is
\begin{equation} 
\langle H \rangle= \langle \frac{p^{2}}{2\mu}\rangle +\langle V\left(r\right)\rangle.
\end{equation}

The strong running coupling constant appeared in the potential V(r) in turn is related to the quark mass parameter as[24,32] 
\begin{equation}
\alpha_{s}\left(\mu^{2}\right)=\frac{4\pi}{\left(11-\frac{2n_{f}}{3}\right)ln\left(\frac{\mu^{2}+M^{2}_{B}}{\Lambda^{2}}\right)}
\end{equation}
where, $n_{f}$ is the number of light flavours[32,33], $\mu$ is the renormalisation scale related to the constituent quark masses as $\mu=2\frac{m_{i}m_{j}}{m_{i}+m_{j}}$ . $M_{B}$ is the background mass related to the confinement term of the potential as $M_{B}=2.24 \times b^{1/2}=0.95 GeV$. The input parameters $\Lambda_{QCD}=0.200GeV$, $b=0.183\;GeV^{2}$ and $cA_{0}$=1$GeV^{2/3}$ are the same with our previous work [24]. \\
Using eq.15 and eq.16 we compute the pseudoscalar ground state masses of the heavy light pseudoscalar mesons and compare with the experimental data in Table.1. The results are found to be in good agreement with the experimental data. Again, using these computed masses we employ eq.14 to obtain the pseudoscalar decay constants. The results are then compared with the available experimental and theoretical values in Table.2. The results are found to be compatible with available experimental and theoretical values.

We note that the present result, ExchQm [34] and that from LC [35] give $f_{B_{S}}>260$ MeV, while other results give $f_{B_{S}}\le 240$
MeV. Hence, the experimental measurements for $f_{B_{S}}$ can be a good testing ground for theoretical reliability of each model
as shown here.
%\newpage 
\begin{table}
\begin{center}
\begin{tabular}{|c| c| c| }\hline
mesons & this work & experimental masses\cite{15}\\\hline%\cline{3-4} 
$D(c\bar{u}/c\bar{d})$&1870.82&$1869.6\pm0.16$ \\\hline
$D(c\bar{s})$&1966.62&$1968\pm 0.33$\\\hline
$B_{u}(\bar{b}u)$&5273.50&$5279\pm 0.29$\\\hline
$B_{s}(\bar{b}s)$&5365.99&$5366\pm 0.6$\\\hline
\end{tabular}
\caption{ Masses of heavy-light  mesons in this work with $m_{d}=0.336 GeV$, $m_{s}=0.465 GeV$, $m_{c}=1.55 GeV$, $m_{b}=4.97 GeV$ and comparison with experimental data. All values are in units of MeV.}
\end{center}
\end{table}
\begin{table}
\begin{center}
\begin{tabular}{c||c|c|c|c|c|c}\hline
&$f_D$&$f_{D_s}$&$f_B$&$f_{B_s}$    \\
\hline
\hline
This work      &$205.14$            &$241.84$             &$201.09$           &$292.04$                            \\
\hline
Experiment[35,36]&$206\pm8.9$         &$260.0\pm5.4$        &$204\pm31$         &$\hspace{0.5cm}\cdots\hspace{0.5cm}$ \\
LQCD  [37]   &$218.9\pm11.3$          &$260.1\pm10.8$      &$196.9\pm8.9$          &$242\pm9.5$                 \\ 
LQCD  [16]    &$213\pm4$          &$248\pm2.5$         &$\hspace{0.5cm}\cdots\hspace{0.5cm}$&$\hspace{0.5cm}\cdots\hspace{0.5cm}$ \\
ExChQm[34]    &$207.53$            &$262.56$             &$208.13$           &$262.39$                             \\
LC    [35]    &$206\pm8.9$        &$267.4\pm17.9$       &$204\pm31$          &$281\pm54$                           \\
LQM   [38]    &$211$               &$248$                &$189$              &$234$                                 \\ 
FC    [39]    &$210\pm10$          &$260\pm10$           &$182\pm8$          &$216\pm8$                             \\
BS    [40,41]    &$230\pm25$          &$248\pm27$           &$196\pm29$         &$216\pm32$                             \\
RQM   [30]    &$234$               &$268$                &$189$              &$218$                                  \\
RPM   [21]    &$208\pm21$          &$256\pm26$           &$198\pm14$          &$237\pm17$                             \\\hline
\end{tabular}
\end{center}
\caption{Decay constants of pseudoscalar heavy-light mesons(in MeV) computed in this work and comparison with experimental[35,36] and theoretical results from (2+1)-flavour asqdat action[37],HPQCD[16], extended chiral quark model(ExChQm)[34],Light cone wavefunction[35],light-front quark model (LQM)[38], field-correlator method (FC)[39], Bethe-Salpeter method (BS)[40,41], relativistic quark model (RQM)[30],relativistic potential model(RPM)[21]}
\end{table}

\subsection{ Leptonic decay width and Branching ratio of $D$, $D_{s}$ and $B$ mesons}
Charged mesons formed from a quark and anti-quark can decay to a charged lepton pair when these objects annihilate via a virtual $W^{\pm}$ boson. Quark-antiquark annihilations via a virtual $W^{+}(W^{-})$ to the $l^{+}\nu(l^{-}\nu)$ final states occur for the $ D^{\pm}$ and $B^{\pm}$ mesons. Purely leptonic decay processes are rare but they have clear experimental signatures due to the presence of a highly energetic lepton in the final state. The theoretical predictions are very clean due to the absence of hadrons in the final state. 
The partial decay width for $P\to\ell\nu$ reads:
\begin{equation}
\label{eq:GAMMALEP}
\Gamma(P\to\ell\nu)=\frac{G^2_F}{8\pi}f^2_P\, M_P\, m^2_\ell\,
\left(1-\frac{m^2_\ell}{M_P} \right)^2|V_{fg}|^2,
\end{equation}

%\begin{equation}
%\Gamma \left(D^{+}_{q}\rightarrow{l^{+}\nu}\right)= \frac{ G^{2}_{F}|V_{cq}|^{2}f^{2}_{D_{q}}}{8\pi}m^{2}_{l}\left(1-\frac{m^{2}_{l}}{M^{2}_{D_{q}}}\right)M_{D_{q}},   q=d,s
%\end{equation}
%and for pure leptonic decay of B mesons
%\begin{equation}
%\Gamma \left(B^{+}\rightarrow{l^{+}\nu}\right)= \frac{ G^{2}_{F}|V_{ub}|^{2}f^{2}_{B}}{8\pi}m^{2}_{l}\left(1-\frac{m^{2}_{l}}{M^{2}_{B}}\right)M_{B}
%\end{equation}

where $G_F$, $P$, $f_P$, $M_P$, and $m_\ell$ denote the Fermi constant, generic pseudoscalar(PS)meson, PS-meson weak-decay constant, PS-meson mass and lepton mass respectively. $V_{fg}$ stands for the CKM matrix element for the quark flavors $f$ and $g$. The importance of measuring $\Gamma \left(P\rightarrow{l\nu}\right)$ depends on the particle being
considered. In the case of the $B^{-}$ meson, the measurement of $\Gamma \left(B^{-}\rightarrow{\tau^{-}\nu}\right)$ provides an indirect
determination of $|V_{ub}|$ provided $f_{B}$ is given by theory.
 
The leptonic widths of the charged PS mesons are obtained by using eq.18 and employing the predicted values of the pseudoscalar masses and decay constants $f_{D}$, $f_{D_{s}}$ and $f_{B}$ . The leptonic widths for separate lepton channels by the choice of $m_{l=\tau,\mu,e}$ are computed to obtain the branching ratio of $D$ and $B$ mesons. The branching ratio of the heavy-light mesons are calculated by using the relation
\begin{equation}
\mathcal{B}=\tau_P \Gamma(P\to\ell\nu).
\end{equation}
The life time of these mesons are $\tau_{D}=1.04 ps$, $\tau_{D_{s}}=0.5 ps$, $\tau_{B}=1.63 ps$ and the CKM elements $V_{cd}=0.230$, $V_{cs}=1.023$, $V_{ub}=3.89\times10^{-3}$ are taken from the world average value reported by Particle data group[15]. The present results as tabulated in Table.3 are in accordance with the available experimental values.
\begin{table}
\begin{center}

\begin{tabular}{|c|c|c|c|c|c|}\hline
mesons              &$ BR_{\tau}\times 10^{-3}$ &$ BR_{\mu}\times 10^{-4}$ &$ BR_{e}\times 10^{-6}$\\\hline
 
$D(cd)$             &1.08 [this work]          &3.89 [this work]               & 0.092[this work]             \\

Expt.[15]           &$< 1.2$                 &$3.82\pm0.32\pm0.09$     & $<8.8$                     \\
B. Patel etal.,[21] &0.9                     &6.6                      & 0.015                   \\\hline
                   &$ BR_{\tau} \times 10^{-2}$ &$ BR_{\mu}\times 10^{-3}$ &$ BR_{e}\times 10^{-4}$\\\hline
$D(cs)$            &5.43 [this work]           &5.33 [this work]           & 0.0013[this work]              \\
HFAG[42]           &$5.38\pm0.32$            &$5.8\pm0.43$             &                \\        
%CLEO-c[36]         &8.0                      &5.94                     &                 \\
%BELLE[37]          &                         &6.44                     &                 \\
Expt.[15]          &$5.6\pm0.4$              &$5.8\pm0.4$              & $<1.2$                 \\
B.Patel etal.,[26] &8.4                       &7.7                     &0.0018                 \\\hline
                   &$ BR_{\tau} \times 10^{-4}$ &$ BR_{\mu}\times 10^{-6}$&$ BR_{e}\times 10^{-6}$ \\\hline
$B(bu)$            &1.07 [this work]            &0.48 [this work]         & 0.0001[this work]                    \\
Wolfgang etal.,[43]& $0.80\pm0.12$            &                       &                     \\
Expt.[10]          &$1.8\pm0.5$               &$<1.0$                 & $<1.9$                \\\hline
\end{tabular}
\caption{Leptonic branching ratio of D, $D_{s}$ and $B$ mesons for three leptonic  channels and comparison with experiment and theoretical results.} 
\end{center}
\end{table}

\section{Summary and Conclusion}
In this work, we have computed the Pseudoscalar masses and decay constants of heavy-light mesons(B and D). We have transformed the wavefunction from $r$ space into momentum space and have used it to obtain the weak decay constants with its relativistic effect. Instead of using a complicated Hamiltonian including a number of terms describing the relativistic corrections and recoil effects etc. we use the simple Hamiltonian and obtain the wave function considering the linear part of the potential $V(r)$ as perturbation.The method of perturbation depends on finding an appropriate 'parent Hamiltonian',for which no general procedure is available, even though the choice may be crucial to the success of the method[28]. As long as the most probable distribution of the wave function in coordinate space is not too large, then treating the linear term of the potential as perturbation seems to be reasonable. Moreover, ZHAO Gong-Bo etal[44] showed that the linear part $br$ of the Cornell Potential can be treated as perturbation in the FLZ(Friedberg, Lee and Zhao) method in an advantageous way with $\alpha_{eff}=4\alpha_{s}/3 \sim 0.3 \sim 0.5$ . In that context, we use stronger $\alpha_{eff}$ in the range of $4\alpha_{s}/3 = 0.50$ to $ \sim 0.55$ to calculate the spin average masses and decay constants. It is  to be noted that the condition of convergence of the Model is being discussed in ref.[22,25], which demands that linear part of the potential can be considered as perturbation provided
\begin{equation}
\frac{(4-\epsilon)(3-\epsilon)\mu b a^{3}_{0}}{2(1+a^{2}_{0}Q^{2})}<<C^{\prime} .
\end{equation}
  The values of the $\alpha_{s}$, used in the computation also follows this condition correctly.\\ 
The computed masses and decay constants are then used to compute the branching ratio of $D$, $D_{s}$ and $B$ mesons for the three leptonic channels $\tau$, $\mu$ and $e$. $B_{s}$ meson being neutral in nature, does not show leptonic decay and hence the leptonic branching ratio for $B_{s}$ meson is not computed. The result of the manuscript is summarised as below.\\
\begin{itemize}
\item The renormalization scale of the model was set to be $\Lambda_{QCD}=200$ MeV, with the approximation that the $\Lambda_{QCD}$ for the heavy-quark effective mass is the same as that for the light quark, taking into account that the QCD-vacuum structure is not affected much by the heavy quarks, although the heavy sources may distort the vacuum to a certain extent. In this energy scale  even the mass  of the light quark in the heavy-light mesons is greater than $\Lambda_{QCD}$ and hence the non-relativistic treatment of the Hamiltonian  is considered to be consistent here. However, if the momentum of the light quark is too larger than the light quark mass, then the
non-relativistic treatment of the energetic operator of the Hamiltonian will not be consistent.

\item The ground state masses of $D$ and $B$ mesons computed in this approach are found to be well consistent with the experimental vaues.
\item  We obtain the decay constants as $f_{D,D_s,B,B_s}=(205.14,\,241.84,\,201.09,\,292.04)$ MeV which are qualitatively compatible with available experimental and theoretical values. Except $f_{B_{S}}$, other values of the decay constants  locate inside the experimental errors. However, with a variation of $\Lambda_{QCD}$ for $D$ and $B$ mesons one obtains more compatible results with the data, although we do not provide those numerical numbers.

\item The computed value in the present work $\frac{f_{D_{s}}}{f_{D}}=1.178$ is found to be in good agreement within the error limit of the recent result Lattice(HPQCD) $\frac{f_{D_{s}}}{f_{D}}=1.164\pm0.018$[16] and Lattice(FNAL and MILC)$\frac{f_{D_{s}}}{f_{D}}=1.188\pm0.025$[45]. However the result of $\frac{f_{B_{s}}}{f_{B}}=1.45$ are found to be larger than the other theoretical values.     
\item The leptonic branching ratio calculated in the present work for three leptonic channels are comparable with their empirical and PDG average data. The large experimental uncertainity in the electron channel makes it difficult for any reasonable conclusion. Furthermore, the ratio of branching ratio in the present work is found to be 
$
R\equiv \frac{{\cal{B}}(D_s^+\to \tau^+\nu)}{{\cal{B}}(D_s^+\to \mu^+\nu)} =10.18
$ which is not far from the experimental result $9.2\pm 0.7$ and Standard Model result 9.76[35].

\end{itemize}

Taking into account all the results summarized above, we can conclude that the present theoretical framework of Potential Model is a qualitatively successful model to study the heavy-light Pseudoscalar mesons. From a phenomenological point of view, the present theoretical framework is a considerably useful tool to investigate various physical quantities for the heavy-light quark systems, such as the Isgur-Wise function, heavy-light meson coupling constants,form factors and Charge radii and so on. Such works are under progress and will appear elsewhere.

\appendix

\numberwithin{equation}{section}
\begin{center}
\section{Appendix}
\end{center}

The Coulomb plus linear potential is given by 
\begin{equation}
V(r)= -\frac{4 \alpha_{S}}{3r}+br+c
\end{equation}

The first order perturbed eigenfunction $\psi^{(1)}$ and first order eigenenergy $W^{(1)}$ using quantum mechanical perturbation theory (Dalgarno's method) can be obtained using the relation 
\begin{equation}
H_{0} \psi^{(1)}+H^{\prime} \psi^{(0)}=W^{(0)} \psi^{(1)}+W^{(1)} \psi^{(0)},
\end{equation}
where 
\begin{equation}
W^{(1)}=< \psi^{(0)} | H^{\prime} | \psi^{(0)} > .
\end{equation}
and 
\begin{equation}
H^{\prime}=br+c
\end{equation}  
Then from ($A.2$), 
\begin{equation} 
(H_{0} - W^{(0)}) \psi^{(1)}=(W^{(1)} - H^{\prime}) \psi^{(0)},
\end{equation}
Putting
\begin{equation}
A=\frac{4 \alpha_{S}}{3},
\end{equation}
we obtain
\begin{equation} 
H_{0}=-\frac{\hbar^{2}}{2 r}\nabla^{2}-\frac{A}{r},
\end{equation}
\begin{eqnarray}
W^{(0)}&=&-\frac{\mu A^{2}}{2} \nonumber\\
&=&-\frac{8\mu \alpha_{S}^{2}}{9}
\end{eqnarray}
and
\begin{eqnarray}
\psi^{(0)}&=&\frac{1}{\sqrt{\pi}}(\mu A)^{\frac{3}{2}} e^{-\mu A r}\nonumber \\
&=&\frac{1}{\sqrt{\pi a_{0}^{3}}} e^{-\frac{r}{a_{0}^{}}}.
\end{eqnarray}
where $\psi^{(0)}$ is the unperturbed wave function in the zeroth order of perturbation and $a_{0}$ is given by equation .
Also, we put $W=W^{(1)}$ , where
\begin{equation}
W^{(1)}=\int \psi_{100}^{*} H^{\prime} \psi_{100} d\tau
\end{equation}
Then  taking $\hbar^{2}=1$ , equation $(A.5)  =>$    
\begin{eqnarray}
\left(-\frac{1}{2\mu}\nabla^{2}-\frac{A}{r}+\frac{\mu A^{2}}{2}\right)\psi^{(1)}=\left(W-br-c\right)\frac{1}{\sqrt{\pi}}(\mu A)^{\frac{3}{2}} e^{-\mu A r}\nonumber \\
\rightarrow\left(\nabla^{2}+\frac{2 \mu A}{r}-\mu^{2}A^{2}\right)\psi^{(1)}=\frac{2\mu}{\sqrt{\pi}}(\mu A)^{\frac{3}{2}} \left(br+c-W\right)
e^{-\mu A r}\nonumber \\
\rightarrow \left(\nabla^{2}+\frac{2}{a_{0}r}-\frac{1}{a_{0}^{2}}\right)\psi^{(1)}=\frac{2\mu}{\sqrt{\pi a_{0}^{3}}} \left(br+c-W\right)
e^{-\frac{r}{a_{0}^{}}}
\end{eqnarray}
Let
\begin{equation}
\psi^{(1)}=(br+c)R(r) 
\end{equation}
\begin{equation}
(A.11)\Longrightarrow \left(\frac{d^{2}}{dr^{2}}+\frac{2}{r}\frac{d}{dr}+\frac{2}{a_{0}^{}r}-\frac{1}{a_{0}^{2}}\right)(br+c)R(r)= D (br+c-W)e^{-\frac{r}{a_{0}}}
\end{equation}
where we put
\begin{equation}
D=\frac{2\mu}{\sqrt{\pi a_{0}^{3}}}. 
\end{equation}
Now,
\begin{equation}
\frac{d}{dr}(br+c)R(r)=br+(br+c)\frac{dR}{dr}
\end{equation}
\begin{equation}
\frac{d^{2}}{dr^{2}}(br+c)R(r)=(br+c)\frac{d^{2}R}{dr^{2}}+2b\frac{dR}{dr}
\end{equation}
Using $(A.15)$ and $(A.16)$ in $(A.13)$, we obtain
\begin{eqnarray}
(br+c)\frac{d^{2}R}{dr^{2}}+2b\frac{dR}{dr}+\frac{2bR}{r}+\frac{2}{r}(br+c)\frac{dR}{dr}+\frac{2}{a_{0}^{}}(br+c)R(r)-\frac{1}{a_{0}^{2}}(br+c)R(r)\nonumber\\
=D (br+c-W)e^{-\frac{r}{a_{0}^{}}}
\end{eqnarray}
Putting 
\begin{equation}
R(r)=F(r)e^{-\frac{r}{a_{0}^{}}}
\end{equation}
\begin{equation}
\frac{dR}{dr}=F^{\prime}e^{-\frac{r}{a_{0}^{}}}-\frac{1}{a_{0}}F(r)e^{-\frac{r}{a_{0}^{}}}
\end{equation}
\begin{equation}
\frac{d^{2}R}{dr^{2}}=F^{\prime \prime}(r)e^{-\frac{r}{a_{0}^{}}}-\frac{2}{a_{0}}F^{\prime}(r)e^{-\frac{r}{a_{0}^{}}}+\frac{1}{a_{0}^{2}}F(r)e^{-\frac{r}{a_{0}^{}}}
\end{equation}
\begin{eqnarray}
(A.17)\Longrightarrow(br+c)\left\{F^{\prime \prime}(r)-\frac{2}{a_{0}^{}}F^{\prime}(r)+\frac{1}{a_{0}^{2}}F(r)\right\}+2b\left\{F^{\prime}(r)-\frac{1}{a_{0}^{}}F(r)\right\}\nonumber \\
+\frac{2b}{r}F(r)+\frac{2}{r}(br+c)\left\{F^{\prime}(r)-\frac{1}{a_{0}^{}}F(r)\right\}+\frac{2}{a_{0}^{}r}(br+c)F(r)\nonumber \\
-\frac{1}{a_{0}^{2}}(br+c)F(r)=D(br+c-W)
\end{eqnarray}
\begin{eqnarray}
\Longrightarrow(br+c)F^{\prime \prime}(r)+\left\{2b+\frac{2}{r}(br+c)-\frac{2}{a_{0}^{}}(br+c)\right\}F^{\prime}(r)\nonumber \\
+\left(\frac{2b}{r}-\frac{2b}{a_{0}^{}}\right)F(r)=D(br+c-W)
\end{eqnarray} 
Let 
\begin{equation}
F(r)=\sum_{n=0}^{\infty} A_{n} r^{n}
\end{equation}
Then,
\begin{equation}
F^{\prime}(r)=\sum_{n=0}^{\infty} n A_{n} r^{n-1}
\end{equation}
and
\begin{equation}
F^{\prime\prime}(r)=\sum_{n=0}^{\infty} n (n-1) A_{n} r^{n-2}
\end{equation}
\begin{eqnarray}
(A.22)\Longrightarrow(br+c)\sum_{n=0}^{\infty} n (n-1) A_{n} r^{n-2}+\left\{2b+\frac{2}{r}(br+c)\right.
\nonumber \\
\left.-\frac{2}{a_{0}^{}}(br+c)\right\}\sum_{n=0}^{\infty} n A_{n} r^{n-1}+\left(\frac{2b}{r}-\frac{2b}{a_{0}^{}}\right)\sum_{n=0}^{\infty} A_{n} r^{n}
=D(br+c-W)
\end{eqnarray} 
\begin{eqnarray}
\Longrightarrow \left\{c\sum_{n=0}^{\infty} n (n-1) A_{n}+ 2c\sum_{n=0}^{\infty} n A_{n}\right\} r^{n-2}+\left\{b\sum_{n=0}^{\infty} n (n-1) A_{n}+4b\sum_{n=0}^{\infty}n A_{n}-\right.
\nonumber \\
\left.\frac{2c}{a_{0}^{}}\sum_{n=0}^{\infty}n A_{n}+2b\sum_{n=0}^{\infty} A_{n}\right\}r^{n-1}+\left(-\frac{2b}{a_{0}^{}}\sum_{n=0}^{\infty} n A_{n} -\frac{2b}{a_{0}^{}}\sum_{n=0}^{\infty}  A_{n}\right) r^{n}=D(br+c-W)
\end{eqnarray} 
Equating coefficients of $r^{-1}$ on both sides of the above identity $(A.27)$,
\begin{equation}
2cA_{1}+2bA_{0}=0
\end{equation}
\begin{equation}
\Longrightarrow (cA_{1}+bA_{0})=0
\end{equation}
Equating coefficients of $r^{0}$ on both sides of the identity $(A.27)$,
\begin{equation}
2cA_{2}+4cA_{2}+4bA_{1}-\frac{2c}{a_{0}^{}}A_{1}+2bA_{1}-\frac{2b}{a_{0}}A_{0}=D(c-W)
\end{equation}
\begin{equation}
\Longrightarrow 6\left(cA_{2}+bA_{1}\right)-\frac{2}{a_{0}}\left(cA_{1}+bA_{0}\right)=D(c-W)
\end{equation}
\begin{equation}
\Longrightarrow cA_{2}+bA_{1}=\frac{1}{6}D(c-W)
\end{equation}
Equating coefficients of $r^{1}$ on both sides of the identity $(A.27)$,
\begin{equation}
12cA_{3}+12bA_{2}-\frac{4c}{a_{0}}A_{2}-\frac{4b}{a_{0}}A_{1}=Db
\end{equation}
\begin{equation}
\Longrightarrow 12(cA_{3}+bA_{2})-\frac{4}{a_{0}^{}}(cA_{2}+bA_{1})=Db
\end{equation}
Using $(A.32)$,
\begin{equation}
12(cA_{3}+bA_{2})-\frac{2}{3 a_{0}^{}}D(c-W)=Db
\end{equation} 
\begin{equation}
\Longrightarrow cA_{3}+bA_{2}=\frac{D}{12}\left\{b+\frac{2}{3a_{0}^{}}D(c-W)\right\}
\end{equation}
Equating coefficients of $r^{2}$ on both sides of the identity $(A.27)$,
\begin{equation}
20(cA_{4}+bA_{3})-\frac{6}{a_{0}^{}}(cA_{3}+bA_{2})=0
\end{equation}
Using $(A.36)$,
\begin{equation}
cA_{4}+bA_{3}=\frac{D}{120a_{0}}\left\{b+\frac{2}{3 a_{0}^{}}D(c-W)\right\}
\end{equation}
From $(A.23)$,
\begin{equation}
F(r)=A_{0}r^{0}+A_{1}r^{1}+A_{2}r^{2}+A_{3}r^{3}+A_{4}r^{4}+\cdots
\end{equation}
Also, from  $(A.12)$ and $(A.18)$,
\begin{eqnarray}
\psi^{(1)}=(br+c)F(r) e^{-\frac{r}{a_{0}^{}}} \nonumber \\
=(br+c)(A_{0}r^{0}+A_{1}r^{1}+A_{2}r^{2}+A_{3}r^{3}+A_{4}r^{4}+\cdots)e^{-\frac{r}{a_{0}^{}}} \nonumber \\
=\left\{cA_{0}r^{0}+\left(cA_{1}+bA_{0}\right)r^{1}+\left(cA_{2}+bA_{1}\right)r^{2}+\left(cA_{3}+bA_{2}\right)r^{3} \right.\nonumber \\
\left.+\left(cA_{4}+bA_{3}\right)r^{4}+\cdots\right\} e^{-\frac{r}{a_{0}^{}}}
\end{eqnarray}
Applying $(A.29)$, $(A.32)$, $(A.36)$ and $(A.38)$ to $(A.40)$
\begin{eqnarray}
\psi^{(1)}=\left[cA_{0}+\frac{1}{6}D(c-W)r^{2}+\frac{D}{12}\left\{b+\frac{2}{3a_{0}^{}}D(c-W)\right\}r^{3} \right. \nonumber \\
\left.+\frac{D}{120a_{0}}\left\{b+\frac{2}{3 a_{0}^{}}D(c-W)\right\}+\cdots\right]e^{-\frac{r}{a_{0}^{}}} 
\end{eqnarray}
Again, from $(A.10)$,
\begin{eqnarray}
W=\int \psi_{100}^{*} (br+c) \psi_{100} d\tau \nonumber \\
=\frac{1}{\pi a_{0}^{3}} \int_{0}^{\infty}(br+c)e^{-\frac{2r}{a_{0}^{}}}r^{2}dr \int_{0}^{\pi} \sin \theta d \theta \int_{0}^{2\pi} d\phi \nonumber \\
=\frac{4}{a_{0}^{3}}\left[b \int_{0}^{\infty}r^{3}e^{-\frac{2r}{a_{0}^{}}}dr+c \int_{0}^{\infty} r^{2} e^{-\frac{2r}{a_{0}^{}}}dr \right]\nonumber \\
=\frac{4}{a_{0}^{3}}\left[b\frac{6 a_{0}^{4}}{16}+c\frac{2a_{0}^{3}}{8}\right] \nonumber \\
=\frac{3}{2}b a_{0}+c  
\end{eqnarray} 
Hence
\begin{equation}
b+\frac{2}{3a_{0}^{}}D(c-W)=0 
\end{equation}
Therefore, $(A.41)$ reduces to 
\begin{eqnarray}
\psi^{(1)}=\left\{cA_{0}+\frac{1}{6}D(c-\frac{3}{2}b a_{0} - c) r^{2}\right\}e^{-\frac{r}{a_{0}^{}}} \nonumber \\
=\left\{cA_{0}-\frac{1}{4}Dba_{0}^{}r^{2}\right\}e^{-\frac{r}{a_{0}^{}}}\nonumber \\
=\left\{cA_{0}-\frac{\mu b a_{0}^{}}{2\sqrt{\pi a_{0}^{3}}}r^{2}\right\}e^{-\frac{r}{a_{0}^{}}}
\end{eqnarray}
where $A_{0}$ is an undeterminable factor appearing in the series solution of the differential equation. Hence, the total wave function correct upto first order of perturbation using Dalgarno's method is given by
\begin{equation}
\psi_{1}^{(1)}=\psi^{(0)}+\psi^{(1)} \nonumber \\
\end{equation}
where $\psi^{(0)}$ and $\psi^{(1)}$ are given by equations $(A.9)$ and  $(A.44)$ respectively. Therefore,
\begin{equation}
\psi_{1}^{(1)}(r)=\frac{1}{\sqrt{\pi a_{0}^{3}}}\left\{1+cA_{0}\sqrt{\pi a_{0}^{3}}-\frac{1}{2}\mu b a_{0}^{}r^{2}\right\} e^{-\frac{r}{a_{0}^{}}}
\end{equation}
Putting
\begin{equation}
C^{\prime}(c)=1+ c A_{0}{\sqrt{\pi a_{0}^{3}}}
\end{equation}
we have finally
\begin{equation}
\psi_{1}^{(1)}\left(r\right)=\frac{1}{\sqrt{\pi a_{0}^{3}}}\left(C^{\prime}(c)-\frac{1}{2}\mu ba_{0}^{}r^{2}\right)e^{-\frac{r}{a_{0}^{}}}
\end{equation}
\end{document}